\documentclass[11pt, a4paper]{article}

\usepackage[utf8]{inputenc}
\usepackage[T1]{fontenc}
\usepackage{lmodern}
\usepackage{amsmath}
\usepackage{amssymb}
\usepackage[margin=2.5cm]{geometry}
\usepackage{booktabs}
\usepackage{graphicx}
\usepackage{float}
\usepackage{array}
\usepackage{microtype}
\usepackage{setspace}
\usepackage{abstract}
\usepackage{titlesec}
\usepackage{authblk}
\usepackage{hyperref}
\usepackage[style=authoryear, backend=biber, natbib=true]{biblatex}

\usepackage{lineno}

\usepackage{pgfplots}
\pgfplotsset{compat=1.17}

\titleformat{\section}{\large\bfseries}{\thesection}{1em}{}
\titleformat{\subsection}{\normalsize\bfseries}{\thesubsection}{1em}{}
\titleformat{\subsubsection}{\normalsize\itshape}{\thesubsubsection}{1em}{}

\title{\textbf{What Do EROIs Measure? Implications for Energy Transition Assessment}}

\author[1]{Thomas Norway}
\author[2]{Olivier Cavalié}

\affil[1]{Independent Researcher, Brussels, Belgium}
\affil[2]{Aix-Marseille Université, CNRS, IRD, INRAE, Collège de France, CEREGE, Aix-en-Provence, France}

\date{}

\begin{document}

\maketitle

\begin{abstract}
Multiple formulations of the Energy Return on Investment (EROI) coexist in the literature, differing mainly in their treatment of self-consumption and external direct energy inputs. This article shows that these differences are not merely conventional: they determine whether EROI measures the net energy surplus available to society or the internal conversion efficiency of the production process. By benchmarking three established formulations against theoretical limit cases, we demonstrate that only the external variant (EXT), which excludes self-consumed energy from the investment term, correctly measures the net energy surplus available to society. The internal (INT) and standard (STD) variants instead converge toward measures of process efficiency. We further show that, in multi-source systems exchanging direct energy flows, consistency requires converting these flows into their upstream embodied energy equivalents before including them in the energy investment term. A generic reallocation formula is derived and shown to uniquely preserve aggregation consistency. Applied to U.S.\ oil and gas production (1919--2007) and China's fossil fuel sectors (1995--2010), the revised framework yields systematically higher EROI values than standard formulations, potentially placing the same physical system on opposite sides of the net energy cliff, with direct implications for energy transition assessment and policy.
\end{abstract}


\bigskip

%\linenumbers

\section{Introduction}

Energy is the fundamental constraint and driver of all 
dissipative structures, including human societies 
\citep{odum2007, roddier2012, prigogine2018}. The 
long-run prosperity of a society depends on its 
capacity to maintain a sufficient energy 
surplus---the net energy remaining after the costs of 
energy production itself have been met 
\citep{hall1986, cleveland1984}. The Energy Return on 
Investment (EROI) metric was developed precisely to 
quantify this surplus, linking the biophysical 
characteristics of energy sources to the 
socio-economic complexity they can sustain 
\citep{hall1986, hall2009}. As the global energy 
transition accelerates, EROI has become a central 
input to assessments of whether low-carbon energy 
systems can deliver the surplus required to support 
complex modern societies \citep{murphy2011, 
murphy2022, delannoy2024, brockway2019}.

Yet multiple formulations of EROI coexist in the 
literature, each reflecting different assumptions 
about what constitutes an energy investment. This 
diversity introduces a problem that is rarely 
acknowledged: reported EROI values for the same 
physical system can differ by a factor of two or 
more depending on the formulation chosen, and 
aggregated and disaggregated assessments of the same 
system may yield inconsistent results. As 
\citet{hall2025} recently noted, such inconsistencies 
in boundary definitions can alter 
reported EROI values and the conclusions drawn from 
them.

The core of the problem lies in the treatment of two 
types of energy flows. First, internal 
self-consumption---energy consumed within the 
production process that never reaches 
society---is treated by most formulations as an 
investment rather than a production loss. Second, 
external direct energy inputs are incorporated at 
face value rather than at their upstream embodied 
energy cost, ignoring that these flows have already 
incurred indirect costs elsewhere in the production 
chain. Both practices conflate energy products with 
their production costs, inflating the measured 
investment and undermining cross-source comparisons.

These inconsistencies are not merely technical. 
Because the relationship between EROI and the net 
energy surplus available to society is strongly 
non-linear---a phenomenon known as the net energy 
cliff \citep{murphy2011_cliff, lambert2014}---even 
moderate distortions in reported EROI can shift the 
implied diagnosis from gradual decline to imminent 
crisis. The choice of EROI formulation thus carries 
direct normative implications for how we evaluate the 
energy transition and the socio-economic trajectories 
it enables.

This article adopts a clear methodological stance: 
the key metric for assessing an energy system's 
contribution to society is the net energy it delivers, 
not the efficiency of its production process. This 
view follows the biophysical tradition in ecological 
economics \citep{georgescu1975, hall1986, odum2007, 
hagens2020}, supported by empirical evidence linking 
energy surplus to socio-economic development 
\citep[e.g.,][]{lambert2014, king2015}.

Building on this perspective, we develop a framework 
for EROI calculation that remains physically 
consistent across system configurations. We evaluate 
three established formulations against theoretical 
limit cases and interconnected multi-source systems, 
and derive a generic reallocation formula for 
interdependent energy-producing units. The framework 
is then applied to two empirical cases---U.S.\ oil 
and gas production over 1919--2007 
\citep{guilford2011} and China's conventional fossil 
fuel sectors over 1995--2010 
\citep{hu2013}---demonstrating that the choice of 
formulation can alter both the magnitude of reported 
EROI and the severity of the implied energy 
constraint.

\section{Biophysical Framework and EROI Formulations}

\subsection{Energy system representation}

To ensure a consistent accounting framework, we model 
energy-producing systems as dissipative 
structures---open thermodynamic systems that maintain 
their organization by processing energy flows 
\citep{prigogine2018, odum2007}. These systems require 
two functionally distinct types of energy inputs: 
(1)~direct energy---energy carriers (such as fuels or 
electricity) consumed in operational processes; and 
(2)~indirect energy---embodied in the goods, 
infrastructure, and services required for the system's 
construction, maintenance, and renewal. The long-term 
viability of the system depends on generating an energy 
surplus beyond these internal requirements.

For instance, in the petroleum supply chain, direct 
energy is consumed when fuel is burned to power 
extraction equipment, while indirect energy is embedded 
in the production of that equipment---the refineries, 
drilling rigs, and transport vehicles that make 
extraction possible.

To formalize this distinction, we group the productive 
activities of the system into three functional blocks:

\begin{itemize}
    \item \textbf{DEP} (Direct Energy Production and 
    distribution): activities that generate energy in a 
    form directly consumable by other parts of the 
    system---such as fuels or electricity---including 
    the capture, conversion, and transportation 
    processes necessary to deliver this energy to its 
    point of use.

    \item \textbf{IEP} (Indirect Energy Production and 
    distribution): activities that produce the 
    resources, equipment, and infrastructure consumed 
    by DEP and other blocks. Because these outputs 
    require energy for their production, the energy 
    embedded in them is referred to as \emph{indirect} 
    (or embodied) energy.

    \item \textbf{DEV} (Development): 
    activities sustained by the surplus energy 
    delivered by the system. This possible surplus may support 
    economic growth (expanding existing productive 
    capacity), social development (healthcare, 
    education, leisure), or technological advancement 
    (increasing the complexity of the productive 
    apparatus)---or any combination thereof.
\end{itemize}

It is important to note that DEP, IEP, and DEV are functional categories, not physical entities. A single productive activity may contribute to more than one block: its inputs and outputs are then allocated proportionally to the function they serve.

\begin{figure}[H]
    \centering
    \includegraphics[width=1\textwidth]{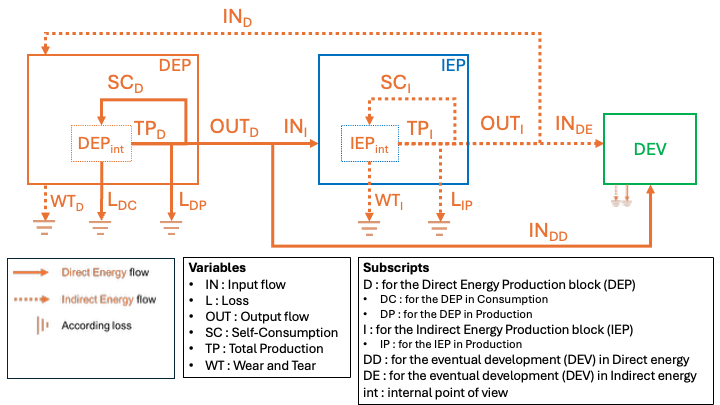}
    \caption{Energy system representation detailing 
    both external and internal flows. The diagram 
    distinguishes between the external boundaries 
    (solid rectangles) of each subsystem (DEP, IEP, 
    DEV) and their internal structures 
    (\textit{DEP\textsubscript{int}} and 
    \textit{IEP\textsubscript{int}}, dashed 
    rectangles). This dual-layer representation allows 
    for a consistent identification of all energy 
    flows necessary for formalizing energy balance 
    equations.}
    \label{fig:final_scheme}
\end{figure}

Fig.~\ref{fig:final_scheme} presents the energy flow 
diagram that serves as the analytical basis for this 
article. It results from three successive 
simplifications of the general schematic of a 
dissipative structure: (i)~external resources are 
internalized within their respective production blocks; 
(ii)~outputs are reclassified as direct or indirect 
energy flows; and (iii)~indirect flows are converted 
into their embodied (``grey'') energy equivalent---the 
cumulative direct energy consumed across all stages of 
their production---rendering them thermodynamically 
comparable to direct flows while implicitly accounting 
for internal IEP losses.

Two features of this representation are essential for 
the analysis that follows. First, the diagram 
distinguishes between external boundaries (solid 
rectangles) and internal structures (dashed rectangles, 
DEP\textsubscript{int} and IEP\textsubscript{int}), 
enabling the identification of both self-consumption 
flows and net external outputs. Second, because 
embodied energy flows are thermodynamically equivalent 
to direct energy flows, both can be compared on a 
common basis---a property that underpins the EROI 
formulations derived in the following section. For 
visual clarity, embodied energy flows are nonetheless 
represented by dashed arrows to distinguish them from 
direct inputs.

\subsection{EROI: Formulations, scope, and significance}

The Energy Return on Investment (EROI) quantifies the 
energy surplus that a production system delivers to 
society. \citet{hall2018} define it as:
\begin{equation}
    \text{EROI} = \frac{\text{Energy returned to society}}
    {\text{Energy required to get and deliver that energy}}
\end{equation}

Two independent choices shape any EROI calculation: 
the \emph{scope} of the energy costs included, and the 
\emph{formula} used to treat internal flows. Both must 
be specified for a metric to be meaningful and 
comparable.

\subsubsection{Scope}
The scope defines which energy costs enter the 
denominator. Three levels are commonly distinguished. 
At the extraction level, only the direct and indirect 
inputs at the production site are counted. At the 
point-of-use level (PoU), the costs of conversion and 
delivery to the consumer are additionally included. At 
the broadest level---the societal EROI 
(EROI\textsubscript{SOC})---the denominator 
encompasses the full supply chain: not only extraction, 
conversion, and delivery, but also the energy embodied 
in infrastructure, equipment manufacturing, labour, 
and all supporting economic activities 
\citep{hall2014}. Societal EROI values are therefore 
substantially lower than their point-of-use 
counterparts. \citet{lambert2014} have shown that 
societal EROI and human development are positively 
correlated, and that countries with a high level of 
development (HDI above 0.75) are generally 
characterized by a societal EROI above approximately 
20:1, although estimates of the minimum threshold for 
a prosperous society remain speculative 
\citep{dupont2021potential}.

This article operates at the point-of-use level. The 
question of scope is therefore held constant 
throughout; what varies between the formulations 
compared below is the treatment of internal energy 
flows.

\subsubsection{Formulations}
\begin{figure}[H]
    \centering
    \includegraphics[width=0.8\textwidth]{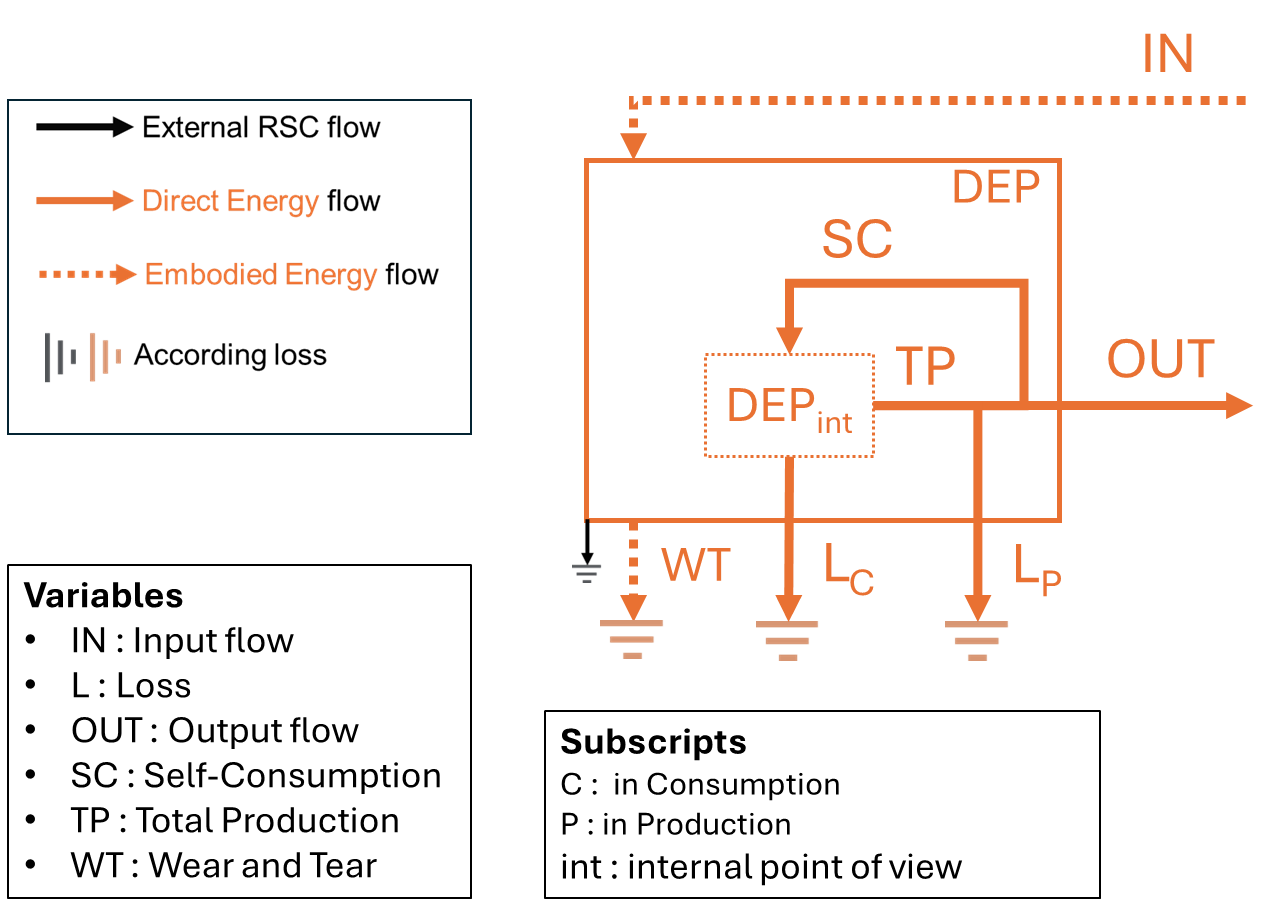}
    \caption{Internal structure of the energy 
    production block (DEP) adapted from 
    Fig.~\ref{fig:final_scheme}. The diagram 
    highlights the main internal losses and 
    self-consumption flows relevant for the 
    formulation of EROI\textsubscript{PoU}.}
    \label{fig:final_scheme_DEP}
\end{figure}

To ensure continuity with the energy system 
representation introduced in the previous section, the 
energy production block (DEP) is extracted from the 
detailed diagram (Fig.~\ref{fig:final_scheme}) and 
redrawn in Fig.~\ref{fig:final_scheme_DEP} with the 
variable definitions relevant to EROI calculation.

Three main formulations of EROI at point of use 
(EROI\textsubscript{PoU}) can be derived from the same 
systemic representation, depending on how 
self-consumption ($SC$) is treated:

\begin{itemize}
    \item \textbf{Internal EROI} (INT) 
    \citep[e.g.,][]{murphy2022}: self-consumption 
    appears in both numerator and denominator, 
    reflecting the total energy flow managed by the 
    system:
    \begin{equation}
       \text{INT} =  \frac{OUT + SC}{IN + SC}
    \end{equation}

    \item \textbf{Standard EROI} (STD) 
    \citep[e.g.,][]{delannoy2024}: self-consumption 
    appears only in the denominator, reflecting the 
    net energy delivered externally while still 
    accounting for internal energy use:
    \begin{equation}
         \text{STD} = \frac{OUT}{IN + SC}
    \end{equation}

    \item \textbf{External EROI} (EXT) 
    \citep[e.g.,][]{aramendia2024}: self-consumption 
    is treated entirely as a production loss and 
    excluded from the denominator:
    \begin{equation}
        \text{EXT} =  \frac{OUT}{IN}
    \end{equation}
\end{itemize}

In all three formulations, production losses ($L_P$) 
are subtracted from the output---they reduce the 
numerator. Self-consumption ($SC$), however, is 
entirely internalized within the system and ultimately 
manifests as conversion losses ($L_C$). From this 
perspective, $SC$ could be treated symmetrically with 
$L_P$---as energy not available for external 
use---which is precisely the logic underpinning the 
EXT formulation.

Hybrid formulations also exist. \citet{aramendia2024}, 
for example, follow an internal-style approach but 
treat final energy self-consumed in transport and 
distribution as a loss rather than an investment. 
This article does not consider hybrid formulations; 
only the three versions defined above (INT, STD, EXT) 
are used in the remainder of the analysis.

\subsubsection{The net energy cliff}
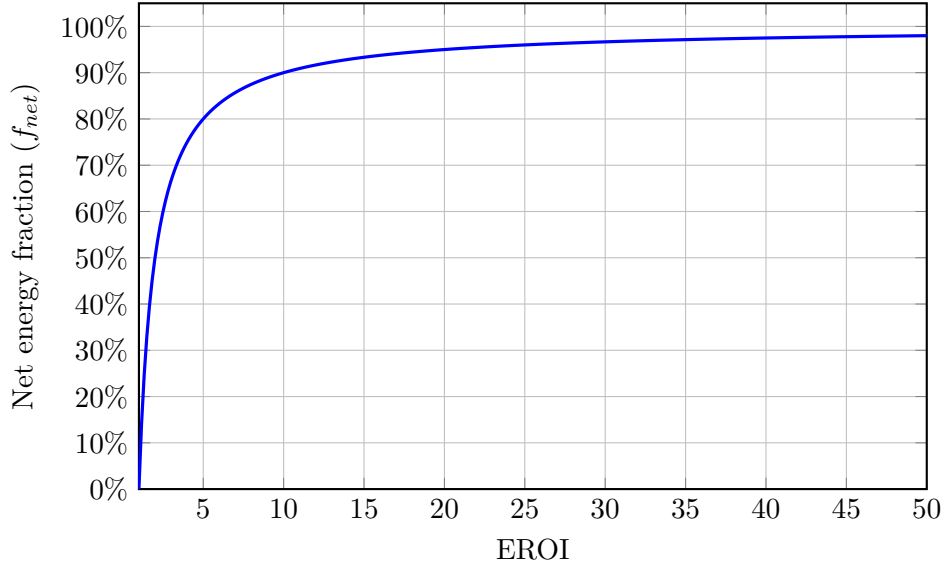
\begin{figure}[ht]
\centering
\begin{tikzpicture}
\begin{axis}[
    width=12cm,
    height=8cm,
    xlabel={EROI},
    ylabel={Net energy fraction ($f_{net}$)},
    xmin=1, xmax=50,
    ymin=0, ymax=1.05,
    grid=major,
    thick,
    ytick={0,0.1,0.2,0.3,0.4,0.5,0.6,0.7,0.8,0.9,1.0},
    yticklabels={0\%,10\%,20\%,30\%,40\%,50\%,60\%,
    70\%,80\%,90\%,100\%},
]

\addplot[blue, very thick, domain=1:50, samples=300] 
    {(x-1)/x};

\end{axis}
\end{tikzpicture}
\caption{The net energy cliff. The curve shows the 
fraction of gross energy available for non-energy 
socio-economic functions as a function of EROI. Below 
approximately 10--15:1, the net fraction declines 
steeply: small reductions in EROI translate into 
disproportionately large losses of available surplus.}
\label{fig:cliff}
\end{figure}

The choice between these formulations carries 
consequences that extend beyond the numerical value 
of the ratio. The fraction of gross energy that 
remains available to society for non-energy purposes 
is:
\begin{equation}
    f_{\text{net}} = \frac{\text{EROI} - 1}{\text{EROI}}
\end{equation}
This relationship is strongly non-linear 
(Fig.~\ref{fig:cliff}): at high EROI values, the net 
fraction is insensitive to moderate declines, but 
below approximately 10--15:1 it drops 
sharply---a phenomenon widely documented as the 
\emph{net energy cliff} \citep{murphy2011_cliff, 
lambert2014, raugei2016, abbott2023}.

Critically, the cliff operates at the societal level: 
it is EROI\textsubscript{SOC}---not the 
point-of-use EROI---that determines the surplus 
effectively available for non-energy economic and 
social functions. Since societal EROI values are 
substantially lower than their point-of-use 
counterparts, formulations that diverge by a factor 
of two or more at the point-of-use level may 
translate into qualitatively different positions on 
the net energy curve once full societal costs are 
accounted for---with direct implications for how 
the viability of energy systems is assessed.

The following section examines which of the three 
formulations yields physically consistent results 
under these criteria.

\subsection{Benchmarking and Consistency Testing}

To clarify the physical meaning of each EROI 
formulation, we perform a benchmarking analysis along 
two axes. We first evaluate each metric against 
extreme limit cases to isolate its underlying physical 
significance. We then examine interconnected 
configurations---scenarios where the energy output of 
one system serves as the input for another. A metric 
is considered consistent only if the aggregate EROI of 
the coupled systems can be derived analytically from 
their individual components without residual error. 
These criteria form the evaluative basis for the 
diagnostic and reformulation work that follows.

\section{A Revised EROI Framework: Diagnosis and Proposal}
\subsection{Sensitivity to OUT and the role of self-consumption}

As \citet{hall2025} note, treating self-consumption ($SC$) as an investment
can distort EROI values, particularly in systems with high $SC$. Including it
in the denominator---as in INT and STD formulations---lowers EROI artificially,
even when the net energy delivered to society ($OUT$) is unchanged. This section
examines this distortion systematically by expressing self-consumption as a
fraction $\alpha$ of total production ($SC = \alpha\,TP$, where $TP = OUT + SC$),
and evaluating each EROI formulation under both normal and extreme conditions.
 
Under this assumption, $SC = \alpha\,OUT/(1-\alpha)$ and $OUT = TP(1-\alpha)$,
so the three formulations become:
 
\renewcommand{\arraystretch}{2.2}
\begin{table}[H]
\centering
\resizebox{\textwidth}{!}{%
\begin{tabular}{|>{\raggedright\arraybackslash}m{0.3\textwidth} 
                |>{\centering\arraybackslash}m{0.27\textwidth} 
                |>{\centering\arraybackslash}m{0.27\textwidth} 
                |>{\centering\arraybackslash}m{0.2\textwidth}|}        
\hline
 & \textbf{INT} & \textbf{STD} & \textbf{EXT} \\
\hline
Generic formulation & $\displaystyle \frac{OUT + SC}{IN + SC}$ &
$\displaystyle \frac{OUT}{IN + SC}$ &
$\displaystyle \frac{OUT}{IN}$ \\[1em]
\hline
With $SC = \alpha \, TP$ &
$\displaystyle \frac{OUT}{(1-\alpha)\,IN + \alpha\,OUT}$ &
$\displaystyle \frac{(1-\alpha)\,OUT}{(1-\alpha)\,IN + \alpha\,OUT}$ &
$\displaystyle \frac{OUT}{IN}$ \\
\hline
\end{tabular}
}
\caption{EROI\textsubscript{PoU} formulations under the assumption that
self-consumption is proportional to total production ($SC = \alpha\,OUT/(1-\alpha)$).}
\label{tab:eroi_sensitivity}
\end{table}
 
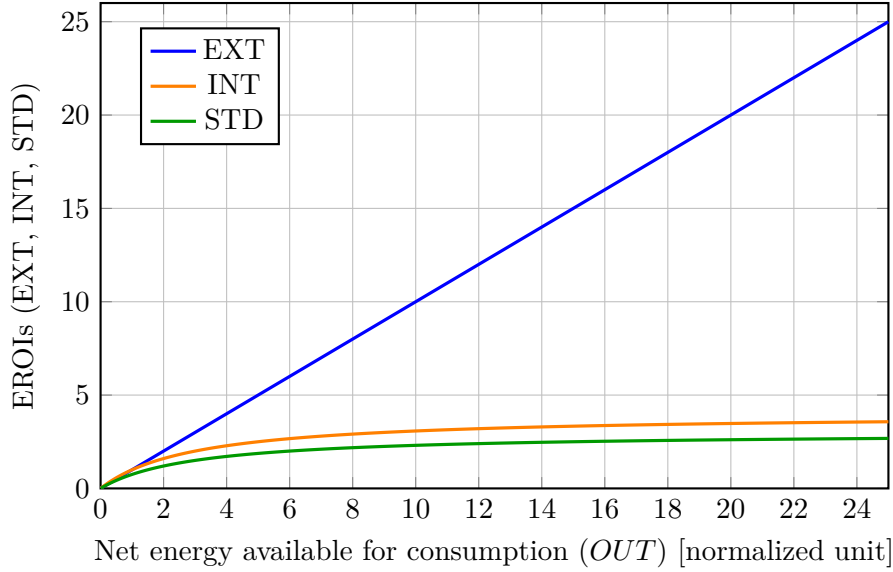
\begin{figure}[ht]
\centering
\begin{tikzpicture}
\begin{axis}[
    width=12cm,
    height=8cm,
    xlabel={Net energy available for consumption ($OUT$) [normalized unit]},
    ylabel={EROIs (EXT, INT, STD)},
    xmin=0, xmax=25,
    ymin=0, ymax=26,
    grid=major,
    legend style={at={(0.05,0.95)},anchor=north west},
    thick
]
% EXT curve: EXT = OUT (with IN=1, alpha=0.25: EXT = OUT)
\addplot[blue, very thick, domain=0.01:25, samples=300] {x};
\addlegendentry{EXT}
% INT curve: INT = TP/(1 + 0.25*TP) = (x/0.75)/(1 + 0.25*x/0.75)
%          = (x/0.75)/(1 + x/3) = (4x/3)/(1 + x/3)
\addplot[orange, very thick, domain=0.01:25, samples=300] {(4*x/3)/(1 + x/3)};
\addlegendentry{INT}
% STD curve: STD = OUT/(1 + 0.25*TP) = x/(1 + x/3)
\addplot[green!60!black, very thick, domain=0.01:25, samples=300] {x/(1 + x/3)};
\addlegendentry{STD}
\end{axis}
\end{tikzpicture}
\caption{Comparison of EROI\textsubscript{PoU} formulations (EXT, INT, STD)
under increasing net energy availability ($OUT$), with $IN = 1$ and
$\alpha = 0.25$. The EXT curve increases linearly with net energy, while INT
and STD converge toward finite asymptotes ($1/\alpha = 4$ and
$(1-\alpha)/\alpha = 3$, respectively) due to the
inclusion of self-consumption in the denominator.}
\label{fig:eroi_methods_curve}
\end{figure}

To visualise the behaviour of these three formulations,
Fig.~\ref{fig:eroi_methods_curve} plots them for a normalized indirect
investment $IN = 1$ and a self-consumption share $\alpha = 0.25$.
 
The divergence is clear: as $OUT$ increases, the INT and STD formulations
saturate toward finite asymptotes determined by $\alpha$ alone, whereas EXT
grows linearly with the net energy actually delivered to the broader system.
In other words, INT and STD become progressively insensitive to increases in
net energy availability---precisely the quantity that an EROI metric should
capture.
 
This methodological concern has not gone unnoticed. \citet{aramendia2024}
acknowledge that excluding self-consumption from the denominator ``may yield
a measure more representative of the potential of the energy system to increase
the net energy supply to society.'' However, they also caution that this
approach ``may yield extremely high values of EROI in the case of highly
self-sufficient systems, overlooking their energy requirements.'' A similar
point is raised by \citet{murphy2022}, who observe that processing energy
inputs often excluded from fossil fuel EROI calculations can change the
resulting values substantially.

These observations are pertinent, but they do not constitute a methodological
flaw. To understand what INT and STD \emph{do} measure, consider the limiting case $IN \to 0$---a thought experiment we refer to as the Horn of Plenty, in which energy is available without any external indirect investment. As previously, the self-consumption is a fraction $\alpha$ of total production. Taking $IN \to 0$, the $OUT$ terms cancel algebraically in the INT and STD expressions (Table~\ref{tab:eroi_sensitivity}), yielding the simplified values in Table~\ref{tab:horn}.

\renewcommand{\arraystretch}{2.2}
\begin{table}[H]
\centering
\begin{tabular}{|>{\raggedright\arraybackslash}m{0.27\textwidth}
                |>{\centering\arraybackslash}m{0.2\textwidth}
                |>{\centering\arraybackslash}m{0.2\textwidth}
                |>{\centering\arraybackslash}m{0.2\textwidth}|}
\hline
 & \textbf{INT} & \textbf{STD} & \textbf{EXT} \\
\hline
$IN \to 0$, $SC = \alpha \, TP$
 & $\displaystyle \frac{1}{\alpha}$
 & $\displaystyle \frac{1-\alpha}{\alpha}$
 & $\displaystyle \infty$ \\
\hline
\end{tabular}
\caption{EROI formulations in a system with zero external investment ($IN \to 0$)
and self-consumption proportional to total production ($SC = \alpha\,TP$, where
$TP = OUT + SC$). The INT and STD values follow from algebraic simplification
of the $OUT$ terms; the EXT formulation diverges as $IN \to 0$.}
\label{tab:horn}
\end{table}

These finite values admit a direct physical interpretation. The internal
conversion efficiency of the DEP block is defined as the fraction of total
production delivered externally:
\begin{equation}
\eta \equiv \frac{OUT}{TP} = \frac{(1-\alpha)\,TP}{TP} = 1 - \alpha
\end{equation}
Substituting $\alpha = 1 - \eta$ into the expressions of Table~\ref{tab:horn}
yields Table~\ref{tab:horn_eta}, which reveals that INT and STD reduce, in this
degenerate case, to functions of $\eta$ alone, independently of the net energy
actually delivered to the rest of the system.

\begin{table}[H]
\renewcommand{\arraystretch}{2.2}
\centering
\begin{tabular}{|>{\centering\arraybackslash}m{0.3\textwidth}
                |>{\centering\arraybackslash}m{0.3\textwidth}
                |>{\centering\arraybackslash}m{0.3\textwidth}|}
\hline
\textbf{INT} & \textbf{STD} & \textbf{EXT} \\
\hline
$\displaystyle \frac{1}{1-\eta}$ &
$\displaystyle \frac{\eta}{1-\eta}$ &
$\displaystyle \infty$ \\
\hline
\end{tabular}
\caption{EROI values as a function of internal conversion efficiency
$\eta = OUT/TP = 1 - \alpha$, in the limiting case $IN \to 0$.}
\label{tab:horn_eta}
\end{table}

This is not a result unique to an extreme case: it is the explicit, analytical expression of the saturation behavior already visible in Figure~\ref{fig:eroi_methods_curve}. INT and STD are mainly determined by
the internal conversion efficiency $\eta$ of the production process,
irrespective of the net energy actually delivered to the rest of the system.
The Horn of Plenty limit simply makes this distinction analytically transparent
by eliminating the external investment term entirely.

This clarifies the two distinct analytical objectives that must not be
conflated. INT and STD are appropriate metrics for assessing the
\emph{internal conversion efficiency} of a production process. EXT is the
appropriate metric for assessing the \emph{net energy contribution to society},
and it is the only formulation that remains sensitive to $OUT$ irrespective of
the level of self-consumption. 
The concern raised by \citet{aramendia2024}---that EXT may ``overlook energy requirements''---applies when the objective is
process diagnostics, not when the objective is societal surplus assessment.
A high EXT value for a highly self-sufficient system 
is therefore not a distortion to be corrected: it 
faithfully reflects a large net surplus, which is 
precisely what the metric is designed to capture.

One might argue that self-consumed energy carries 
an opportunity cost, since it could in principle 
be delivered to society 
\citep[see][]{guilford2011, hall2025}. While this 
reasoning is economically sound, it does not alter 
the biophysical accounting: the energy in question 
has been consumed within the production process 
and is no longer available to the rest of the 
structure. Treating it as an investment on these 
grounds conflates a potential revenue with an 
actual cost---and, as shown above, causes the 
metric to converge toward a measure of internal 
process efficiency rather than societal surplus, 
regardless of the opportunity cost reasoning.

Thus, the following sections focus exclusively
on the EXT metric, examining first how it behaves in systems with multiple
interconnected DEP units, and then how direct energy inputs should be properly
accounted for in its calculation.

\subsection{Embodied Energy Accounting in Multi-Source Systems}

The preceding sections have established that, within the EXT framework, direct self-consumption ($SC$) must be treated as a production loss rather than an energy investment. While this principle resolves the internal accounting of a single DEP unit, it leaves open a broader question: how should EXT be computed in systems composed of several interconnected DEP units that exchange direct energy flows with one another?

This question is not merely hypothetical. Real energy systems are almost never isolated: a coal mine consumes electricity generated from wind farm, an oil refinery uses natural gas in its processes. In all such cases, one DEP unit supplies a direct energy flow that is functionally equivalent to the self-consumption of another. The way this cross-flow is accounted for has direct consequences for the consistency of EROI assessments across system boundaries.

\begin{figure}[H]
    \centering
    \includegraphics[width=0.75\textwidth]{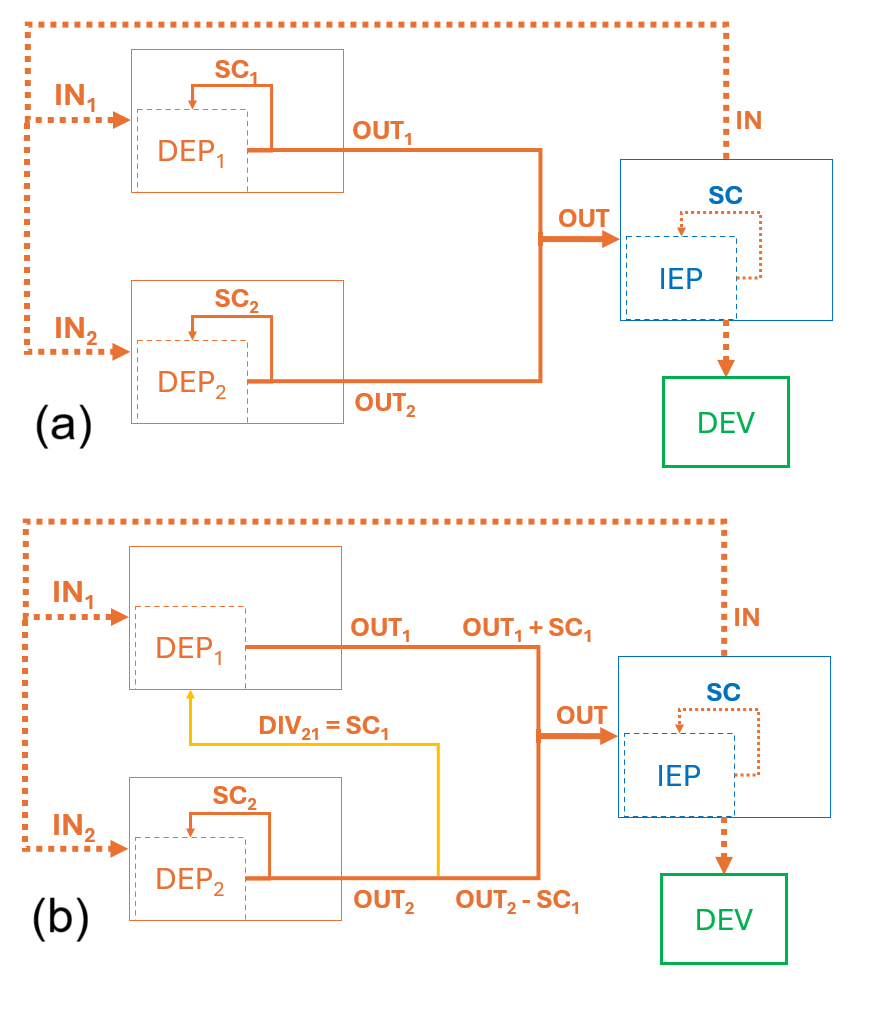}
    \caption{Simplified representation of two DEP units supplying a common IEP block. (a) Configuration A: each DEP unit uses a portion of its own output to operate (self-consumption). (b) Configuration B: a fraction $\text{DIV}_{21} = \text{SC}_1$ of DEP$_2$'s output is diverted to cover DEP$_1$'s self-consumption. The total energy delivered to IEP is identical in both configurations.}
    \label{fig:multi_dep}
\end{figure}

To address this, we consider a minimal two-source system in which a single IEP block is supplied by two DEP units, DEP$_1$ and DEP$_2$ (Fig.~\ref{fig:multi_dep}a). The two units need not be identical in their production rates or indirect investments: the only requirement is that the energy carriers they produce are of the same physical nature, so that one can substitute for the other. Each unit $i$ receives an indirect investment ($\text{IN}_i$), incurs an internal self-consumption ($\text{SC}_i$), and delivers a net output ($\text{OUT}_i$) to IEP.

In this baseline configuration---hereafter Configuration A---the system-wide external EROI can be written as a weighted average of the individual EROIs of each DEP unit, with weights given by their respective indirect investments:

\begin{equation}
\begin{aligned}
\label{eq:eroi_syst}
\text{EROI}_{\text{syst}}^{A} &= \frac{\text{OUT}_{1} + \text{OUT}_{2}}{\text{IN}_1 + \text{IN}_2} \\
&= \frac{\text{IN}_{1}(\text{OUT}_{1}/\text{IN}_1) + \text{IN}_{2}(\text{OUT}_{2}/\text{IN}_2)}{\text{IN}_1 + \text{IN}_2} \\
&= \frac{\text{IN}_{1} \cdot \text{EROI}_1 + \text{IN}_{2} \cdot \text{EROI}_{2}}{\text{IN}_1 + \text{IN}_2}
\end{aligned}
\end{equation}

Because both DEP units produce energy carriers of the same nature, it is physically possible to redirect a fraction of DEP$_2$'s output to cover the self-consumption needs of DEP$_1$, in lieu of DEP$_1$ self-supplying. This alternative arrangement---Configuration B (Fig.~\ref{fig:multi_dep}b)---diverts a flow $\text{DIV}_{21} = \text{SC}_1$ from DEP$_2$ to DEP$_1$. As a result, DEP$_1$ no longer needs to retain $\text{SC}_1$ internally and delivers $\text{OUT}_1 + \text{SC}_1$ to IEP, while DEP$_2$ delivers only $\text{OUT}_2 - \text{SC}_1$. The total energy reaching IEP remains strictly unchanged: $(\text{OUT}_1 + \text{SC}_1) + (\text{OUT}_2 - \text{SC}_1) = \text{OUT}_1 + \text{OUT}_2$.

Two physical principles must therefore hold across the two configurations. First, since the global energy balance is preserved, the system-wide EROI must be identical: $\text{EROI}_{\text{syst}}^{A} = \text{EROI}_{\text{syst}}^{B}$. Second, DEP$_2$ has not changed in any physical sense between A and B: it processes the same indirect investment and produces the same gross output, irrespective of whether its energy is subsequently consumed by IEP or by DEP$_1$. Its individual EROI is therefore invariant: $\text{EROI}_2^{A} = \text{EROI}_2^{B} \equiv \text{EROI}_2$. Together, these two constraints define the consistency criterion that any coherent EROI accounting must satisfy.

A first-pass treatment of Configuration B would leave $\text{IN}_1$ and $\text{IN}_2$ unchanged. The total indirect investment is necessarily conserved between the two configurations:

\begin{equation}
\text{IN}_1^{B} + \text{IN}_2^{B} = \text{IN}_1 + \text{IN}_2
\end{equation}

Fixing $\text{IN}_2^{B} = \text{IN}_2$ therefore forces $\text{IN}_1^{B} = \text{IN}_1$, which in turn determines $\text{EROI}_1^{B} = \text{OUT}_1^{B} / \text{IN}_1^{B}$ by construction. This is incompatible with the physical reality of Configuration B: DEP$_1$ now delivers $\text{OUT}_1 + \text{SC}_1$ to IEP with the same indirect investment, so its EROI must have increased. The consistency criterion therefore admits no solution under the assumption that indirect investments remain individually unchanged.

The resolution is that although the \emph{physical} indirect investments mobilised upstream of each unit are unaltered, their \emph{attribution} to DEP$_1$ and DEP$_2$ must be reallocated to reflect the diversion. Since a fraction of DEP$_2$'s production now sustains DEP$_1$'s operations, a corresponding fraction of the indirect investment $\text{IN}_2$ that was required to produce this diverted flow must be transferred from DEP$_2$'s account to DEP$_1$'s account. Let $x$ denote this transferred quantity. The effective indirect investments in Configuration B become $\text{IN}_1^{B} = \text{IN}_1 + x$ and $\text{IN}_2^{B} = \text{IN}_2 - x$, while the total system-wide indirect investment is preserved: $\text{IN}_1^{B} + \text{IN}_2^{B} = \text{IN}_1 + \text{IN}_2$.

Substituting these reallocated investments into the system-wide EROI for Configuration B and equating with Configuration A yields:

\begin{equation}
\label{eq:eroi_div}
\frac{\text{IN}_1 \dfrac{\text{OUT}_1}{\text{IN}_1} + \text{IN}_2 \cdot \text{EROI}_2}{\text{IN}_1 + \text{IN}_2} = \frac{(\text{IN}_1 + x) \dfrac{\text{OUT}_1 + \text{SC}_1}{\text{IN}_1 + x} + (\text{IN}_2 - x) \cdot \text{EROI}_2}{\text{IN}_1 + \text{IN}_2}
\end{equation}

Simplifying the inner ratios on both sides---where each indirect investment cancels with the denominator of its paired EROI term---and cancelling common elements, this equation reduces to $\text{SC}_1 = x \cdot \text{EROI}_2$, which yields the unique solution:

\begin{equation}
\label{eq:x_solution}
x = \frac{\text{SC}_1}{\text{EROI}_2} = \frac{\text{SC}_1}{\text{OUT}_2} \cdot \text{IN}_2
\end{equation}

The reallocated indirect investment $x$ is therefore exactly equal to the fraction of $\text{IN}_2$ proportional to the share of DEP$_2$'s output that is diverted to DEP$_1$. This result has a transparent physical meaning: the diverted direct energy flow $\text{DIV}_{21}$ carries with it, in effect, its own upstream indirect cost---namely the share of $\text{IN}_2$ that was mobilized to produce it. Two limit cases confirm the consistency of this result. If $\text{SC}_1 = 0$ (no diversion takes place), then $x = 0$: no reallocation is needed, and Configurations A and B are identical. Conversely, if $\text{SC}_1 = \text{OUT}_2$ (the entirety of DEP$_2$'s production is diverted to sustain DEP$_1$), then $x = \text{IN}_2$: the full indirect investment of DEP$_2$ is transferred to DEP$_1$, which is consistent with the fact that DEP$_2$ no longer contributes any net energy to IEP.

This result extends well beyond the two-unit case. The diverted flow $\text{DIV}_{21}$ is a direct energy carrier, yet what enters the indirect-investment denominator of DEP$_1$'s EROI is not $\text{DIV}_{21}$ itself, but its embodied-energy equivalent $x = \text{DIV}_{21} / \text{EROI}_2$. Incorporating $\text{DIV}_{21}$ at face value would violate the homogeneity of the energy balance: the EXT denominator must contain only indirect (grey) energy, and any direct energy flow entering it must first be converted into the corresponding upstream investment that produced it.

Crucially, this is not merely a methodological preference. It is the unique solution imposed by the requirement that individual EROIs aggregate consistently into the system-wide EROI. Any alternative treatment---and in particular the common practice of incorporating direct energy inputs at face value into the EROI denominator---would produce individual EROIs that are mathematically incompatible with the EROI of the aggregated system (see Appendix A for an explicit example). The consistency criterion thus reinforces, on purely formal grounds, the principle that has guided this article from the outset: only indirect (embodied) energy flows constitute genuine energy investments in a coherent EROI accounting framework. The following section generalizes this reallocation rule to systems composed of an arbitrary number of interconnected DEP units.

\subsection{Generic formulation}

The result derived in the preceding section extends 
naturally to systems composed of an arbitrary number 
of interconnected DEP units. The diverted flow 
$\text{DIV}_{21}$ introduced above is generalized 
here as $\text{DIV}_{jk}$, denoting the direct energy 
diverted from any upstream unit DEP$_j$ to a 
downstream unit DEP$_k$. Consider a downstream unit 
DEP$_k$ that receives such flows from $n$ upstream 
units, each $j \in \{1, \ldots, n\}$.

By direct extension of the two-unit derivation, the indirect investment that must be reallocated from DEP$_j$ to DEP$_k$ to account for this diverted flow is the corresponding share of $\text{IN}_j$:

\begin{equation}
\label{eq:div}
\text{IN}_{j \to k} = \frac{\text{DIV}_{jk}}{\text{OUT}_j} \cdot \text{IN}_j = \frac{\text{DIV}_{jk}}{\text{EXT}_j}
\end{equation}

where the second equality follows from the definition $\text{EXT}_j = \text{OUT}_j / \text{IN}_j$. The diverted direct energy flow is thereby expressed in terms of the embodied energy that was required upstream to produce it.

Summing these contributions over all upstream units supplying DEP$_k$ yields the generic expression for its external EROI:

\begin{equation}
\label{eq:generic}
\text{EXT}_k = \frac{\text{OUT}_k}{\text{IN}_k + \displaystyle\sum_{j=1}^{n} \frac{\text{DIV}_{jk}}{\text{EXT}_j}}
\end{equation}

where $\text{OUT}_k$ is the net energy output of DEP$_k$ delivered to downstream blocks (IEP or other DEP units), $\text{IN}_k$ is its own indirect investment, and the summation captures the embodied-energy cost of all direct energy inputs drawn from upstream DEP units.

In the case of mutually interdependent DEPs---where $EXT_k$ appears 
simultaneously in the denominator of several units---Equation~\ref{eq:generic} 
generates a coupled linear system, the simultaneous solution of 
which is detailed in Appendix B.

This formulation preserves the full biophysical consistency of the EXT metric across nested and interconnected configurations: every direct energy exchange between units is accounted for in terms of its upstream indirect cost, ensuring that individual EROIs always aggregate consistently into the system-wide EROI.

A similar reallocation principle has been applied, albeit without explicit justification, by \citet{colla2024}. Their third EROI variant (EROI$_3$) implicitly adopts this logic through input-output analysis. While the resulting implementation appears consistent with the framework developed here, the underlying rationale and boundary assumptions are not elaborated in detail, leaving room for further discussion in a broader comparative analysis.

\section{Applications}

To illustrate the practical implications of the revised framework, we reanalyse two empirical datasets from the literature using the EXT formulation. The first application---U.S.\ oil and gas production---demonstrates the effect of reclassifying direct energy from an investment to a loss in a single aggregated sector. The second---China's conventional fossil fuels---illustrates the use of the generic reallocation formula (Equation~\ref{eq:generic}) in a system with cross-flows between interdependent production units. In both cases, we compare our revised $\text{EROI}_{\text{EXT}}$ with the $\text{EROI}_{\text{STD}}$ reported by the original authors.

\subsection{Application to U.S.\ Oil and Gas Production}
\label{sec:us}

We apply the EXT framework to the dataset compiled by 
\citet{guilford2011} for U.S.\ oil and gas production 
over the period 1919--2007. Their study reports direct 
and indirect energy inputs separately (Table~6 in 
\citealt{guilford2011}), making it well suited to our 
reanalysis. (Note: the direct energy value for 1954 in 
Table~6 of \citealt{guilford2011} appears to contain a 
typographical error---53.9~PJ---that is inconsistent 
with both the reported total of 1096~PJ and the 
detailed appendix data; we use a corrected value of 
902~PJ throughout.) An alternative estimate of indirect 
costs by \citet{guilford2011} (Table~7, based on 
O'Connor and Cleveland's depreciation series) is used 
as a sensitivity check.

Using the notation of \citet{guilford2011}, where 
$E_{\text{out}}$ corresponds to $OUT$, 
$E_{\text{direct}}$ to $SC$, and 
$E_{\text{indirect}}$ to $IN$ in our framework, 
\citet{guilford2011} compute a standard EROI following 
the protocol of \citet{murphy2011}:
\begin{equation}
\text{STD} = \frac{E_{\text{out}}}{E_{\text{direct}} + E_{\text{indirect}}}
\end{equation}
where both direct and indirect energy inputs appear in 
the denominator. In contrast, the EXT formulation 
retains only indirect energy investments in the 
denominator:
\begin{equation}
\text{EXT} = \frac{E_{\text{out}}}{E_{\text{indirect}}}
\end{equation}

Because the U.S.\ oil and gas sector is treated as a single aggregated DEP unit, no inter-unit reallocation is required: the direct energy inputs reported by \citet{guilford2011}---dominated by natural gas consumed on-site for pumping and pressurization---are reclassified entirely as self-consumption. This represents the simplest possible application of the EXT logic.

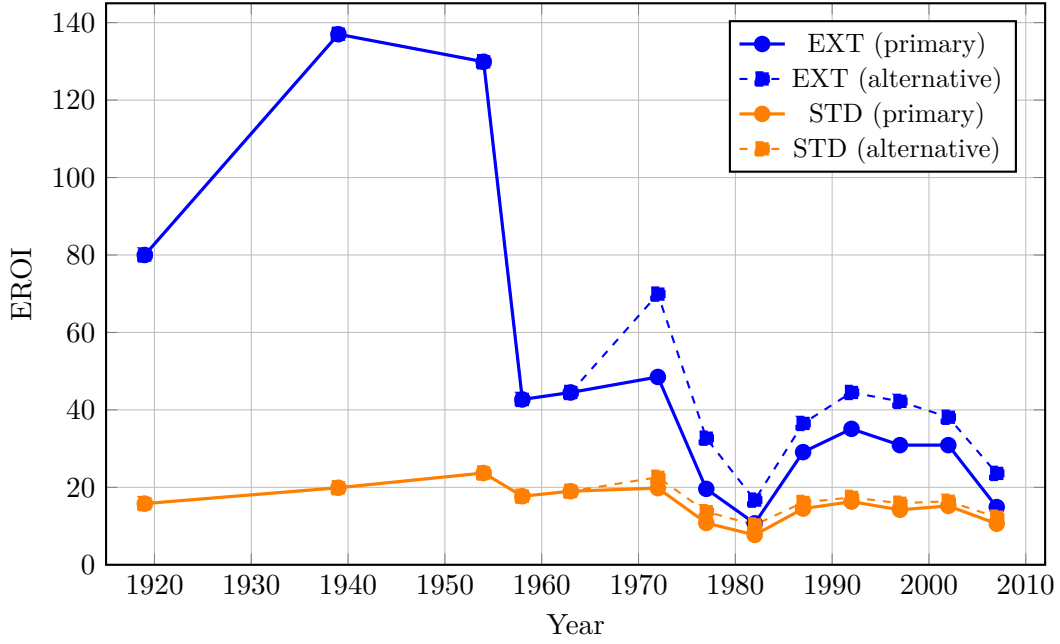
\begin{figure}[ht]
\centering
\begin{tikzpicture}
\begin{axis}[
    width=14cm,
    height=9cm,
    xlabel={Year},
    ylabel={EROI},
    xmin=1915, xmax=2012,
    ymin=0, ymax=145,
    xtick={1920,1930,1940,1950,1960,1970,1980,1990,2000,2010},
    xticklabel style={/pgf/number format/1000 sep={}},
    grid=major,
    legend style={at={(0.97,0.97)}, anchor=north east, font=\small},
    thick,
    every axis plot/.append style={mark size=2.5pt}
]

% EXT - Table 6 (primary)
\addplot[blue, very thick, mark=*] coordinates {
    (1919,80.0) (1939,137.0) (1954,129.9) (1958,42.7) (1963,44.5)
    (1972,48.5) (1977,19.6) (1982,10.7) (1987,29.1) (1992,35.1)
    (1997,30.9) (2002,30.9) (2007,14.9)
};
\addlegendentry{EXT (primary)}

% EXT - Hybrid T6+T7 (alternative)
\addplot[blue, dashed, mark=square*] coordinates {
    (1919,80.0) (1939,137.0) (1954,129.9) (1958,42.7) (1963,44.5)
    (1972,69.9) (1977,32.7) (1982,16.7) (1987,36.5) (1992,44.5)
    (1997,42.2) (2002,38.1) (2007,23.6)
};
\addlegendentry{EXT (alternative)}

% STD - Table 6 (primary)
\addplot[orange, very thick, mark=*] coordinates {
    (1919,15.8) (1939,19.9) (1954,23.7) (1958,17.7) (1963,19.0)
    (1972,19.8) (1977,10.8) (1982,7.7) (1987,14.5) (1992,16.3)
    (1997,14.2) (2002,15.2) (2007,10.6)
};
\addlegendentry{STD (primary)}

% STD - Hybrid T6+T7 (alternative)
\addplot[orange, dashed, mark=square*] coordinates {
    (1919,15.8) (1939,19.9) (1954,23.7) (1958,17.7) (1963,19.0)
    (1972,22.6) (1977,13.7) (1982,10.3) (1987,16.1) (1992,17.4)
    (1997,15.9) (2002,16.4) (2007,12.3)
};
\addlegendentry{STD (alternative)}

\end{axis}
\end{tikzpicture}
\caption{Comparison of Standard EROI (STD) and External EROI (EXT) for U.S.\ oil and gas production, 1919--2007 with corrected 1954 value; see text. Solid lines use primary data from Table~6 of \citet{guilford2011}; dashed lines use alternative indirect cost estimates from Table~7.}
\label{fig:us_eroi}
\end{figure}

\subsubsection{Results and interpretation}
Fig.~\ref{fig:us_eroi} compares STD as 
reported by \citet{guilford2011} with the revised EXT over the full study period. 
Two features stand out. First, EXT values 
are systematically higher than their 
STD counterparts, by a factor that 
ranges from approximately 1.4 in periods of high indirect 
investment (1982, 2007) to over 5 in the early decades when 
indirect costs were comparatively small. This gap reflects the 
reclassification of direct energy---which constitutes between 
30\% and 60\% of total inputs depending on the period---from 
an investment to a production loss.

Second, both formulations confirm the same long-run declining 
trend, consistent with progressive resource depletion. The 
qualitative diagnosis is therefore robust to methodological 
choice. However, the two metrics yield substantially different 
values: STD for 2007 is approximately 
11:1, while EXT ranges from 15 to 
24:1 depending on the indirect cost estimate. As discussed in 
Section~2.2, the net energy cliff operates at the societal 
level, where EROI values are substantially lower than at the 
point of use. A divergence of this magnitude at the 
point-of-use level may therefore translate into qualitatively 
different positions on the net energy curve once full societal 
costs are accounted for, with direct implications for how the 
severity of the energy constraint is assessed.

The sensitivity analysis using alternative indirect cost 
estimates (dashed lines in Fig.~\ref{fig:us_eroi}) confirms 
that this divergence is robust: regardless of which dataset is 
used, the EXT values remain substantially above the STD values 
and the qualitative pattern is preserved.

\subsection{Application to China's Conventional Fossil Fuels}
\label{sec:china}

The Chinese energy system provides a more complex test case, as it involves two interdependent production sectors---oil \& gas (DEP$_1$) and coal (DEP$_2$)---that exchange direct energy flows with one another. This configuration requires the generic reallocation formula (Equation~\ref{eq:generic}) to ensure consistent accounting of upstream embodied energy costs.

We apply the EXT framework to the empirical dataset provided by \citet{hu2013} for the period 1995--2010. Their study distinguishes direct and indirect energy inputs for both sectors, making it possible to identify self-consumption flows, cross-sector diversions, and indirect investments separately. The detailed classification of energy flows and the treatment of electricity inputs are presented in Appendix B; here we summarize the key principles and results.

\begin{figure}[H]
    \centering
    \includegraphics[width=0.8\textwidth]{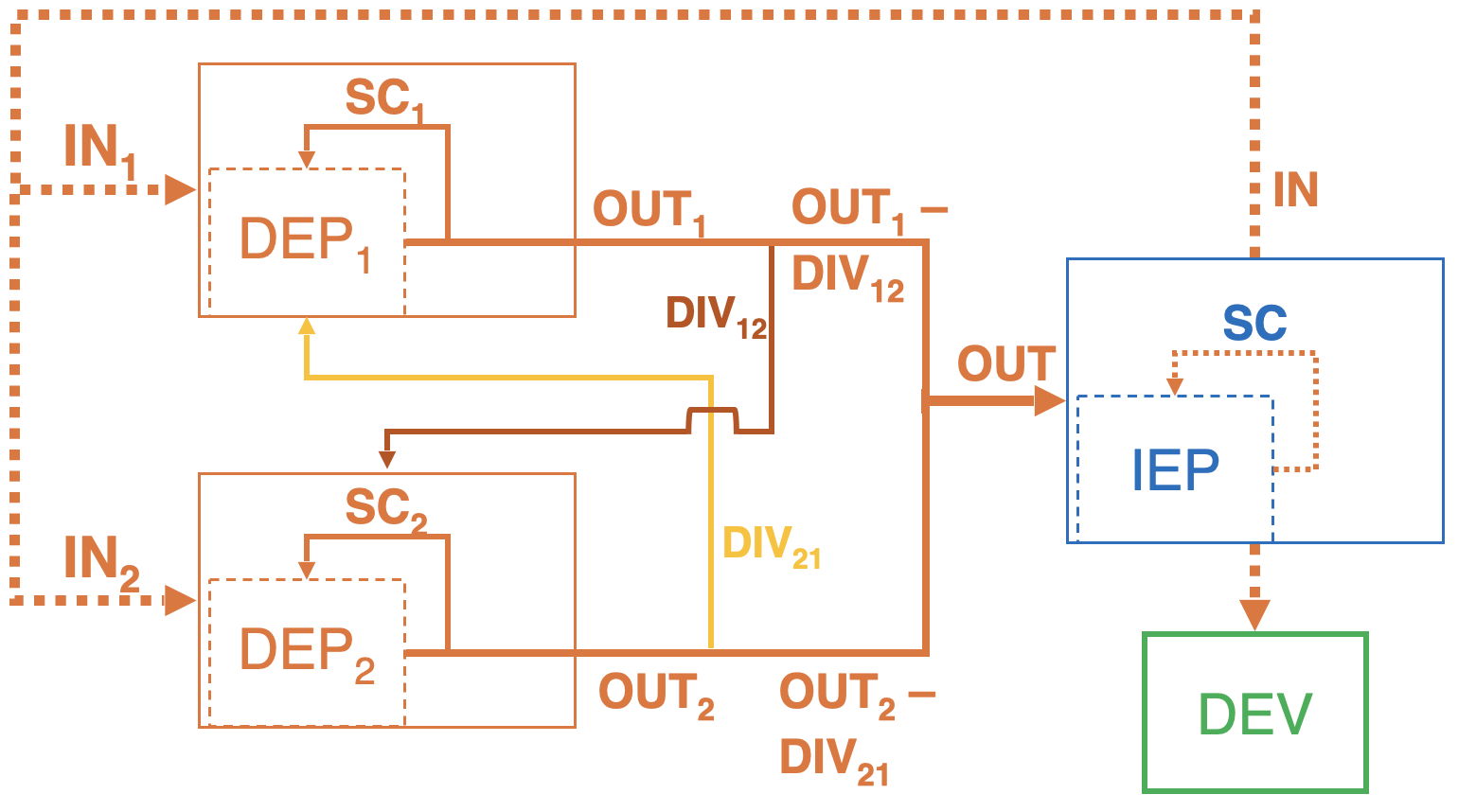}
    \caption{Simplified energy flow configuration for the Chinese energy system. DEP$_1$ and DEP$_2$ represent oil \& gas and coal production units, respectively. A fraction of the gross output from each unit ($OUT_i$) is diverted back into the system as direct self-consumption ($SC_i$) or as cross-input flows ($DIV_{12}$, $DIV_{21}$) to sustain production. The resulting net output ($OUT$) feeds the Indirect Energy Production (IEP) block, which ultimately delivers energy to the economic development sector (DEV). Dotted lines ($IN$) represent the embodied energy flows required to maintain and operate the production infrastructure.}
    \label{fig:china_config}
\end{figure}

Under the EXT framework, direct self-consumption ($SC$) comprises energy carriers of the same type as those produced by the sector, or carriers whose upstream cost can be expressed directly in terms of that sector's own output. The remaining direct inputs---coal and electricity in the oil \& gas sector, diesel and gasoline in the coal sector---originate from external DEP units and are converted into their upstream embodied energy equivalents using Equation~(\ref{eq:generic}) before entering the denominator. Indirect energy inputs, already expressed as embodied energy costs in \citet{hu2013}, enter the denominator unchanged.

\begin{figure}[H]
    \centering
    \includegraphics[width=1\textwidth]{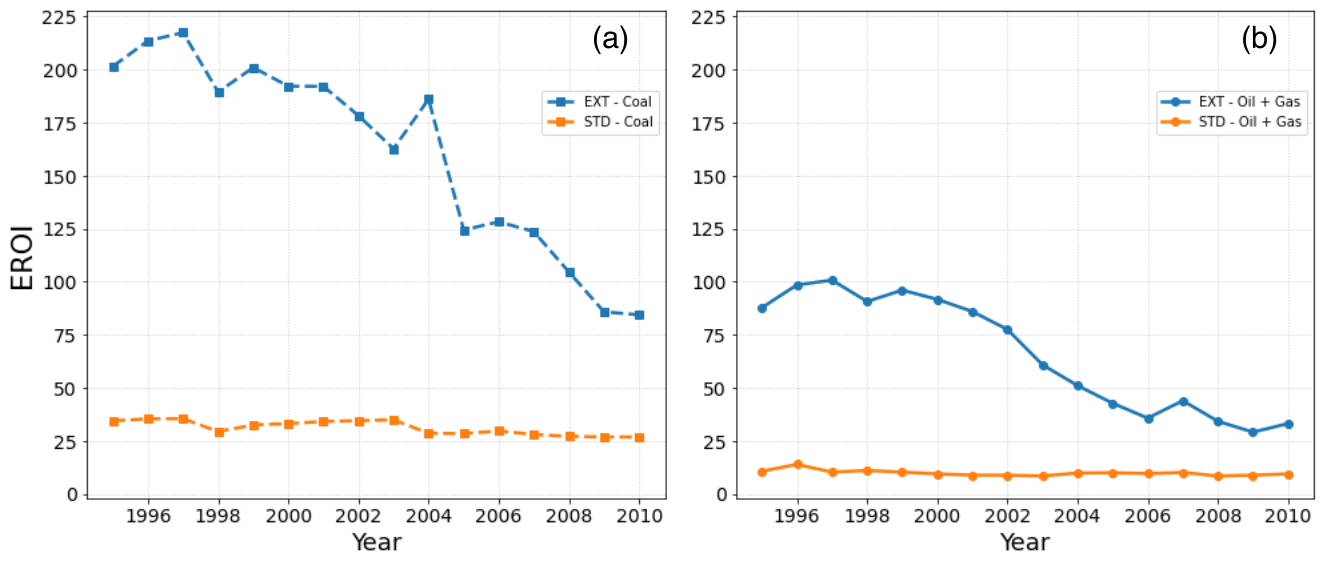}
    \caption{Comparison of External EROI (EXT) and Standard EROI (STD) for Coal (a) and Oil \& Gas (b) sectors in China, 1995--2010.}
    \label{fig:china_eroi}
\end{figure}

\subsubsection{Results and interpretation}
Fig.~\ref{fig:china_eroi} compares STD as reported by \citet{hu2013} with the revised EXT for both sectors. The pattern is consistent with the U.S.\ case: EXT values are systematically higher than their 
STD counterparts, and both formulations confirm a declining trend over the study period. The gap is particularly pronounced for coal, where direct self-consumption 
represents a larger share of gross production. 

As in the U.S.\ application, the two formulations yield substantially different EROI values for the Chinese system: approximately 10--14:1 versus 35--95:1 for oil \& gas, and 25--35:1 versus 85--220:1 for coal (STD and EXT, respectively). These are extraction-level metrics; translating them into societal-level EROI---which accounts for the full upstream energy costs of the goods and services required by the energy sector---substantially reduces the resulting values. Given the strongly non-linear relationship between societal EROI and net energy availability, a divergence of the magnitude observed here at the extraction level may place the same physical system on opposite sides of the net energy cliff, depending on which formulation is used.

Both applications thus illustrate the same core finding: the choice of EROI formulation does not merely rescale the metric---it can determine whether an energy system appears to be approaching a critical constraint or undergoing a gradual decline. This distinction carries direct implications for how energy transition scenarios are evaluated and for the policy 
responses they motivate.

\section{Discussion and Conclusion}

This article has addressed two intertwined questions. 
The first is conceptual: if EROI is to serve as an 
indicator of the net energy surplus that underpins 
societal prosperity and complexity 
\citep{hall1986, hall2018, lambert2014}, then the 
metric must capture the energy actually made available 
to society---not the internal efficiency of 
production processes. The second is methodological: 
what should count as an energy investment in the EROI 
denominator? The analysis shows that these two 
questions have a single answer.

First, benchmarking the three main EROI formulations 
against limit cases reveals that the INT and STD 
variants do not measure the net energy surplus 
available to society. In the limiting case of zero 
external investment, both reduce to functions of the 
DEP block's internal conversion efficiency 
(Section~3.1), irrespective of the energy actually 
delivered to the rest of the system. Only the EXT 
formulation captures this surplus consistently.

Second, when multiple energy-producing units exchange 
direct energy flows, consistency requires that these 
flows be converted into their upstream embodied energy 
equivalent before entering the EROI denominator 
(Section~3.3). The generic reallocation formula 
(Equation~\ref{eq:generic}) ensures that individual 
EROIs aggregate without residual error into the 
system-wide EROI---a property that no alternative 
treatment of direct energy inputs can satisfy.

Applied to U.S.\ oil and gas production 
\citep{guilford2011} and to China's conventional 
fossil fuel sectors \citep{hu2013}, the revised 
framework yields EROI values that are systematically 
higher than those obtained with standard formulations. 
Because the net energy cliff operates at the societal 
level---where EROI values are substantially lower 
than at the point of use---a divergence of this 
magnitude may place the same physical system on 
opposite sides of the critical threshold, depending 
on which formulation is used. The choice of 
methodology is therefore not neutral: it shapes the 
diagnostic and, ultimately, the policy response.

\paragraph{Recommendations}
To reduce the ambiguity that currently surrounds EROI 
reporting, we propose two complementary measures.

First, we suggest clarifying the terminology applied 
to legacy EROI variants to reflect their actual 
analytical scope. As shown in Section~3.1, the INT 
and STD formulations measure the internal conversion 
efficiency of the production process rather than the 
net energy delivered to society. We propose grouping 
them under the term \emph{EROC} (Energy Return on 
Consumption), which reflects their common 
characteristic: measuring how much energy survives 
internal self-consumption. Under this convention, the 
term \emph{EROI} would be reserved exclusively for 
the external formulation, the only metric shown here 
to be physically consistent across system 
configurations, and systematically accompanied by its 
scope: $\text{EROI}_{\text{ext}}$ (extraction), 
$\text{EROI}_{\text{PoU}}$ (point of use), or 
$\text{EROI}_{\text{soc}}$ (societal). EROI values 
computed at different scopes are not directly 
comparable, and omitting this specification is a 
recurring source of confusion in the literature.

Second, we recommend systematic application of the 
generic reallocation formula 
(Equation~\ref{eq:generic}) whenever direct energy is 
exchanged between production subsystems. This ensures 
that the EXT metric remains coherent across nested 
and interconnected configurations---a prerequisite 
for meaningful cross-source and cross-scenario 
comparisons in energy transition assessments.

\paragraph{Limitations and perspectives}
Several limitations should be acknowledged. The 
empirical applications presented here rely on 
secondary data from existing studies, and the 
classification of energy flows into self-consumption 
versus external inputs involves judgement calls that 
may differ across analysts. The reallocation formula 
assumes proportional attribution of indirect costs, 
which may not hold in all real-world configurations.

Additionally, the EROI values reported here are 
computed on an annual basis, yet energy investments 
and their returns are not temporally aligned: 
infrastructure built in one year may produce energy 
for decades. Cumulative EROI formulations, which 
integrate inputs and outputs over the full production 
history, may offer a more representative picture of 
long-term energetic performance, but require 
continuous time-series data that are rarely available.

A further source of uncertainty lies in the 
conversion of indirect monetary costs into energy 
units. Both datasets reanalysed here rely on sectoral 
energy intensity factors (MJ per monetary unit) that 
are themselves subject to methodological debate and 
temporal variation. These two issues---temporal 
aggregation and monetary-to-energy 
conversion---warrant dedicated investigation in 
future work.

Future research should also test the framework 
against primary datasets and extend it to renewable 
energy systems, where the structure of direct and 
indirect inputs differs substantially from fossil 
fuels.

More broadly, EROI---however correctly 
calculated---captures only one dimension of an energy 
system's performance. It does not account for 
externalities that may severely affect planetary 
boundaries, such as greenhouse gas emissions, 
biodiversity loss, or land use. A high EROI does not 
guarantee sustainability: an increasing share of the 
surplus may be needed to mitigate the environmental 
consequences of the production that generated it. 
Conversely, a lower EROI does not disqualify 
technologies that are aligned with ecological and 
social objectives, provided they remain viable within 
the overall energy surplus. Integrating EROI with 
environmental and social criteria remains an essential 
direction for future research in ecological economics.

\printbibliography

%\nolinenumbers

% --- APPENDICES START HERE ---

\newpage
\noindent\Large{\textbf{Appendices}}
\appendix

\normalsize
\section{Illustrative example of Inconsistent EROI Accounting}
\label{sec:appendix_eroi}

This appendix provides a toy numerical example illustrating the inconsistency discussed in Section 3.2: incorporating direct energy inputs at face value into the EROI denominator leads to violations of system-level aggregation consistency.

We consider the same two-DEP (Direct Energy Production) configuration as in Section 3.2 and compare two treatments: 
(i) a \textbf{consistent formulation}, in which diverted direct energy flows are converted into their embodied energy equivalents, and 
(ii) an \textbf{inconsistent formulation}, in which they are included at face value.

To assess the aggregation consistency of each treatment, we evaluate the external $\text{EROI}_{\text{PoU}}$ formulation (EXT) across four distinct system boundaries (Table \ref{tab:eroi_by_scope}):
\begin{enumerate}
    \item The energy performance of $\text{DEP}_1$ in isolation;
    \item The energy performance of $\text{DEP}_2$ in isolation;
    \item The combined performance of both units aggregated into a single subsystem, $\text{DEP}_{\text{T}}$ (DEP Total);
    \item The weighted average performance of $\text{DEP}_1$ and $\text{DEP}_2$ based on their respective energy investments, $\text{DEP}_{\text{A}}$ (DEP Average).
\end{enumerate}

This fourth perimeter allows us to assess whether the aggregation of EROI values weighted by energy investment leads to the same result as computing the EROI on the aggregated energy flows. In other words, we test whether:

\begin{equation}
\text{EROI}_{\text{PoU}}^{\text{A}} = \frac{IN_1 \cdot \text{EROI}_{1} + IN_2 \cdot \text{EROI}_{2}}{IN_1 + IN_2} \quad \stackrel{?}{=} \quad \text{EROI}_{\text{PoU}}^{\text{T}}
\end{equation}

To simplify expressions in the $\text{DEP}_{\text{T}}$ and $\text{DEP}_{\text{A}}$ rows, we introduce the aggregated notations: $OUT_T = OUT_1 + OUT_2$, $IN_T = IN_1 + IN_2$, and $SC_T = SC_1 + SC_2$, denoting the total usable output, indirect investment, and direct self-consumption across both units.

\begin{table}[H]
\renewcommand{\arraystretch}{1.8}
\centering
\begin{tabular}{l|cccc}
%\toprule
 & $\text{DEP}_1$ & $\text{DEP}_2$ & $\text{DEP}_{\text{T}}$ & $\text{DEP}_{\text{A}}$ \\
\midrule
EXT & $\dfrac{OUT_1}{IN_1}$ & $\dfrac{OUT_2}{IN_2}$ & $\dfrac{OUT_T}{IN_T}$ & $\text{EROI}^{\text{A}}$ \\
\bottomrule
\end{tabular}
\caption{System boundaries and corresponding external EROI formulations used for the consistency test.}
\label{tab:eroi_by_scope}
\end{table}

To evaluate internal consistency, we assign numerical values to the parameters. For both $\text{DEP}_1$ and $\text{DEP}_2$, we set $IN_1 = IN_2 = 1$, $OUT_1 = 10$, $OUT_2 = 20$, $SC_1 = 9$, and $SC_2 = 1$. 

\begin{figure}[H]
    \centering
    \includegraphics[width=0.8\textwidth]{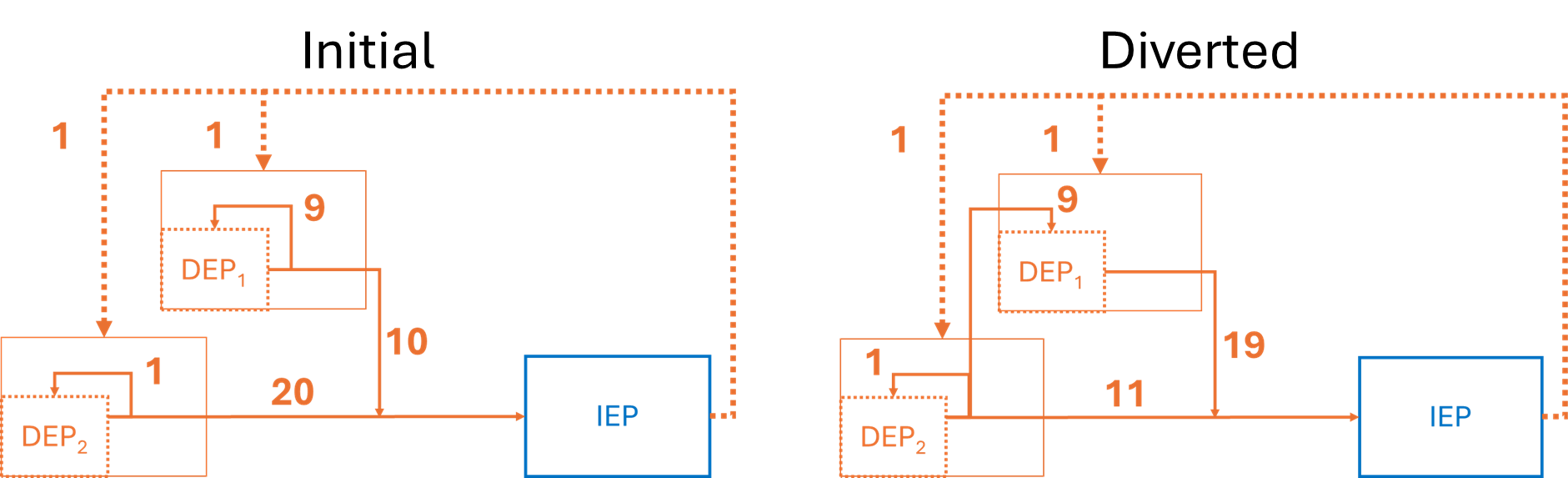}
    \caption{Comparison of initial and diverted configurations in an asymmetric energy production scenario.}
    \label{fig:dep_asymmetry}
\end{figure}

Computations are performed for two configurations (Fig. \ref{fig:dep_asymmetry}): 
(1) the \textbf{Initial} setup, where each DEP operates independently; and 
(2) the \textbf{Diverted} setup, where a portion of $\text{DEP}_2$'s output is redirected to supply $\text{DEP}_1$. Both configurations maintain constant physical energy totals but present structural variations used to assess whether the EROI results remain stable and coherent.

This numerical scenario highlights a critical limitation: despite the physical structure and total energy balances remaining stable, the standard formulation does not provide consistent results across all system boundaries. Once production is structurally reorganized, incorporating diverted direct energy at face value into the denominator causes the average value ($\text{EXT}_{\text{A}}$) to diverge from the total system value ($\text{EXT}_{\text{T}}$).
Table~\ref{tab:numerical_eroi_diversion} illustrates this inconsistency for the EXT formulation.

\begin{table}[H]
\renewcommand{\arraystretch}{1.5}
\centering
\begin{tabular}{lcc}
\toprule
\textbf{System boundary} & \textbf{Initial} & \textbf{Diverted} \\
\midrule
$\text{DEP}_1$ & 10 & 19 \\
$\text{DEP}_2$ & 20 & 20 \\
$\text{DEP}_{\text{T}}$ & 15 & 15 \\
$\text{DEP}_{\text{A}}$ & 15 & \textcolor{red}{\textbf{19.5}} \\
\bottomrule
\end{tabular}
\caption{Numerical values of the external $\text{EROI}_{\text{PoU}}$ formulation for different system boundaries under initial and diverted configurations.}
\label{tab:numerical_eroi_diversion}
\end{table}

Applying the reallocation framework developed in Section~3.2, where diverted direct energy flows are converted into their upstream embodied energy equivalent, we recompute the effective energy investments. Instead of assuming equal weights ($IN_1 = IN_2 = 1$), we apply the allocation rule derived in the main text. This yields adjusted values of $IN_1 = 1.45$ and $IN_2 = 0.55$.

\begin{table}[H]
\renewcommand{\arraystretch}{1.5}
\centering
\begin{tabular}{lcc}
\toprule
\textbf{System boundary} & \textbf{Diverted (face value)} & \textbf{Diverted (embodied)} \\
\midrule
$\text{DEP}_1$ & 19 & 13.1 \\
$\text{DEP}_2$ & 20 & 20 \\
$\text{DEP}_{\text{T}}$ & 15 & 15 \\
$\text{DEP}_{\text{A}}$ & \textcolor{red}{\textbf{19.5}} & \textcolor{green!60!black}{\textbf{15.0}} \\
\bottomrule
\end{tabular}
\caption{Numerical values of the EROI across system boundaries using the consistent reallocation framework.}
\label{tab:numerical_eroi_ext}
\end{table}

Table \ref{tab:numerical_eroi_ext} shows that once diverted direct energy flows are expressed in terms of their upstream embodied energy cost, coherence is restored across all system boundaries.

%%%%%%%%%%%%%%%%%%%%% CHINA %%%%%%%%%%%%%%%%%%%%%%%

\section{China Case Study: Data Sources and Application of the Reallocation Framework}
\label{app:china}

\subsection{Source Data}

All data are drawn from \citet{hu2013}, whose empirical analysis covers
China's oil and gas extraction sector and coal production sector over the
period 1995--2010. The original data originate from the  \citet{csy1995-2010} and the 
\citet{cesy1995-2010}. Tables~A1--A4 of \citet{hu2013} provide raw and
converted energy flows for the oil and gas sector; Tables~B1--B4 cover the
coal production sector.

\subsection{Treatment of Electricity Inputs}
\label{sec:elec}

Over the period 1995--2010, coal accounted for approximately 75--80\% of
China's total electricity generation \citep{iea2012weo}. Electricity inputs
are therefore treated as coal-derived in both sectors, but the accounting
treatment differs depending on which sector receives them.

\paragraph{Coal sector}
Electricity consumed within the coal production sector is reclassified as
self-consumption (\(SC\)). Since it is generated by burning coal in
thermal power plants, it ultimately draws on the sector's own output.
Given a mean thermal efficiency of approximately 33\% for Chinese
coal-fired power plants over the study period \citep{zhao2010}, the
primary coal equivalent of an electricity input \(E_{\text{elec}}\)
(expressed in PJ of final electrical energy) is:
\begin{equation}
    E_{\text{elec}}^{\text{primary}} = \frac{E_{\text{elec}}}{\eta}
    \approx 3 \cdot E_{\text{elec}}
    \label{eq:elec_coal}
\end{equation}
This value is added to the other \(SC\) flows and subtracted from gross
coal output, consistently with the treatment of all other coal-derived
carriers. The approach is thermodynamically explicit: rather than treating
electricity as an external input whose embodied cost must be traced through
the production chain, it directly expresses the upstream coal consumption
implied by that electricity use. 

The thermal efficiency of Chinese coal power plants improved modestly over
the study period, from approximately 32\% in the mid-1990s to around
35--36\% after 2005 \citep{zhao2010}. Using a fixed factor of three
therefore slightly overestimates the primary coal equivalent in later
years, marginally increasing \(SC\) and thus reducing
\(\text{EROI}_{\text{EXT}}\) for the coal sector. This bias is
conservative and quantitatively small.

\paragraph{Oil and gas sector}
Electricity consumed within the oil and gas extraction sector is treated
as an external direct input originating from the coal sector DEP and is added to raw coal to get the total direct diverted energy from coal DEP.

\paragraph{Note on power plant infrastructure}
For both sectors, the conversion of electricity into its primary coal equivalent accounts only for the fuel consumed in generation. The indirect energy embodied in the power plant infrastructure itself is not included, as these costs are not reported in the source data. In principle, a share of these infrastructure costs proportional to the fraction of electricity consumed by each energy production sector should be allocated to its indirect investment ($IN$). This omission marginally overestimates the external EROI of both sectors.

\subsection{Upstream embodied energy of diverted direct energy flows}
\label{app:china:oilgas}

As demonstrated in Section~3.2, the upstream embodied 
energy corresponding to diverted direct energy flows 
must be taken into account to compute the external 
EROI. In the case of China, both sectors exchange 
direct energy flows: the coal sector supplies coal 
and electricity to oil \& gas extraction 
($\text{DIV}_{21}$), while the oil \& gas sector 
supplies diesel and gasoline to coal production 
($\text{DIV}_{12}$). To determine the upstream 
embodied energy associated with these diverted flows, 
the following coupled system is established based on 
Equation 8:

\begin{equation} \label{eq:embodied_energy}
\begin{cases}
IN_{21} = \dfrac{\text{DIV}_{21}}{\text{OUT}_2} 
\cdot (\text{IN}_2 + IN_{12}) \\[1em]
IN_{12} = \dfrac{\text{DIV}_{12}}{\text{OUT}_1} 
\cdot (\text{IN}_1 + IN_{21})
\end{cases}
\end{equation}

where $IN_{jk}$ denotes the upstream embodied energy 
reallocated from sector $j$'s investment budget to 
sector $k$, and $\text{DIV}_{jk}$ denotes the 
corresponding diverted direct energy flow from 
sector $j$ to sector $k$. The subscripts 1 and 2 
refer to oil \& gas and coal, respectively, 
consistently with Fig.~7.

Due to the mutual interdependence between $IN_{12}$ 
and $IN_{21}$, the system constitutes a coupled 
linear system. Since $\text{DIV}_{jk} < \text{OUT}_j$ 
by construction, the determinant is strictly 
positive, guaranteeing a unique solution for 
$(IN_{12},\, IN_{21})$ for each year of the study 
period.

The resulting upstream embodied energy values are 
added to the own indirect investment $\text{IN}_k$ 
of each sector, and the external EROI is then 
computed as:
\begin{equation}
\text{EXT}_k = \frac{\text{OUT}_k}{\text{IN}_k + IN_{jk}}
\end{equation}
for each year, yielding the annual estimates reported 
in Fig.~8.

\end{document}